\documentclass[aps,nopacs,amsmath,amssymb,twocolumn]{revtex4}
\usepackage{graphicx}
\usepackage{dcolumn}
\usepackage{bm}
\usepackage{epsf}

\begin{document}

\title{Molecular fields and statistical field theory of fluids. Application to interface phenomena.
}

\author{Nikolai V.\ Brilliantov}
\affiliation{Skolkovo Institute of Science and Technology (Skoltech), 121205, Moscow, Russia}
\affiliation{Departmentof Mathematics, University of Leicester, Leicester LE1 7RH, UK}

\author{J. Miguel Rub\'{\i}}
\affiliation{Departament de Fisika Fonamental, Universitat de Barcelona, Av.Diagonal 647, 08028 Barcelona, Spain}
\affiliation{PoreLab, Department of Physics, Norwegian University of Science and Technology, Norway}
\author{Yury A.\ Budkov}
\affiliation{Tikhonov Moscow Institute of Electronics and Mathematics, National Research University Higher School of Economics, Tallinskaya st. 34, 123458 Moscow, Russia}
\affiliation{G.A. Krestov Institute of Solution Chemistry, Russian Acad.\ Sci., 153045, Academicheskaya st. 1, Ivanovo, Russia}

\date{\today}
\begin{abstract}
 Using the integral transformation, the  field-theoretical Hamiltonian of the statistical field theory of fluids is obtained, along with the microscopic expressions for the coefficients of the Hamiltonian. Applying this approach to the liquid-vapor interface, we derive an explicit {\em analytical} expression for the surface tension in terms of temperature, density and parameters of inter-molecular potential. We also demonstrate that a clear physical  interpretation may be given to the formal statistical field arising in the integral transformation -- it may be  associated with the one-body local microscopic potential. The results of the theory, lacking any ad-hoc or fitting parameters are in a good agreement with available simulation data.
\end{abstract}

\pacs{68.03.Cd, 68.03.-g, 68.18.Jk, 05.70.Np}

\maketitle

\section{Introduction}
The growing popularity of the field theoretical (FT) methods in statistical physics reflects recognition of the power and flexibility of such methods \cite{Amit:78,Brezin}.  In most of the FT  approaches the configuration integral, associated with a thermodynamic potential (free energy,  Gibbs free energy, etc.) is expressed in terms of a functional integral over one or few space-dependent fluctuating fields, emerging in the Kac-Siegert-Stratonovich-Hubbard-Edwards (KSSHE) transformation \cite{Stratonovich,Kac,Hubbard1,hubbard,edwards1965statistical,Edwards,Siegert}. Commonly, this field for simple fluids is treated as a formal mathematical object facilitating the analysis. Up to our knowledge, the query, whether a physical interpretation to this field may  be given, has not been  risen yet.

Once the functional integral representation is obtained, one can apply  standard field-theoretical techniques to find the configuration integral and space correlation functions. These tools comprise the mean-field (or saddle-point) approximation, e.g.  \cite{Amit:78,Brezin,Caillol3,Caillol2,RussierCaillol2010,Brill:98,IvanLis}, random phase approximation, e.g.  \cite{budkov2019Astatistical,budkov2018nonlocal,budkov2019statistical,zakharov1999classical,advzic2014field}, Gaussian equivalent representation, e.g.  \cite{efimov1996partition,baeurle2002method}, many-loop expansion, e.g.  \cite{netz2001electrostatistics}, variation method, e.g. \cite{lue2006variational}, and renormalization group theory, e.g.  \cite{BrezinFeng:84,brilliantov1998peculiarity}.

The field theoretical methods are successfully applied to describe thermodynamic and structural properties of simple and complex fluids, non-homogeneous fluids and fluid interfaces   and have  already a half-century history \cite{Storer,IvanLis,Caprio,Caillol3,Caillol2,Caillol1,Parola,efimov1996partition,RussierCaillol2010,Frusawa2018,Edwards,edwards1965statistical,hubbard,Brill:98}. In the pioneering paper \cite{Storer} Storer outlined the derivation of the equation of state of simple fluid, treating separately the repulsive (short range) and attractive parts of the interaction potential. He expressed the grand partition function in terms of the functional integral, with the coefficients depending  on the  thermodynamic and structural properties of the reference fluid with the short-range potential. The properties of the reference fluid, such as the equation of state and structure factor, were supposed to be known. The functional integration has been  then performed under the random phase approximation.  The approach was close to the one developed by Edwards \cite{Edwards} for ionic fluids, where the excluded volume interactions between ions were taken into account to improve the  Debye-Hueckel theory.

A similar field theory of simple fluids has been proposed by Hubbard and Schofield \cite{hubbard}. They also divided the total inter-molecular potential into repulsive and attractive parts and recast the grand partition function  into the form of a functional integral \cite{hubbard}. The exponential factor  in the functional integral was written as an effective  magnetic-like Hamiltonian, expressed in terms of functional series of a  fluctuating field $\phi(\bold{r})$. The latter mimics the magnetization field in magnetics. The coefficients of the effective Hamiltonian were, in their turn, written as multi-particle correlation functions of the reference fluid with purely repulsive interactions.

Using this effective Hamiltonian, the authors further discussed, whether the Wilson's  theory  of criticality was applicable to fluid criticality. They demonstrated that the modified RG analysis applied to the magnetic-like Hamiltonian, proved the Ising-like criticality of simple fluids.  Although the main focus of the study \cite{hubbard} was the fluid criticality, the authors also showed  that the coefficients of the field-theoretical Hamiltonian could  be related to the microscopic properties of the reference system. This was in a sharp contrast to the phenomenological theories,  see e.g. \cite{eugene1971introduction,landau2013course,Igna:2000}, where such Hamiltonians, used to analyse the near-critical behavior of fluids and interface phenomena had phenomenological coefficients.

The derivation of the effective field theoretical Hamiltonian  has been completed in  \cite{Brill:98}. Here all the coefficients have been found and  explicitly expressed in terms of the thermodynamic and structural characteristics of the reference hard-core fluid, namely, in terms of its compressibility and zero moments of multi-particle correlation functions. The microscopic expression for the Gizburg criterion \cite{landau2013course} for fluid criticality has been also reported \cite{Brill:98}. Somewhat alternative approaches for the field-theoretical description of simple fluids and liquid-vapor  interface have been developed in Refs.  \cite{IvanLis,Caillol1,Caillol3,RussierCaillol2010}. Although the microscopic expressions for the coefficients of the field-theoretical Hamiltonain could  be, in principle, obtained in such approaches, this was beyond the scope of the above studies; the  physical nature of the field was not also addressed.

As it has been already mentioned, the KSSHE integral transformation yields the Hamiltonian that depends on the
statistical field, which mimics the magnetization field in magnetics \cite{Brill:98}. The magnetic-like  form of the
Hamiltonian is very convenient to  analyze critical and interface phenomena \cite{BrezinFeng:84,Igna:2000}. In
particular, one can find an equilibrium space distribution of the magnetization with an interface. Finding then the
free energy per unit area of the interface, one obtains the surface tension. Still, this purely phenomenological
approach does not provide surface tension in terms of molecular parameters, but rather  the expressions in terms of the
phenomenological coefficients of the magnetic-like Hamiltonian \cite{landau2013course}. It seems also interesting to
find a possible physical interpretation of the formal field in the field-theoretical Hamiltonian.

In the present study we provide the microscopic, molecular expressions for the parameters of the magnetic-like field theoretical Hamiltonian and reveal the physical nature of the stochastic field exploited in the field theories of fluids. Using these microscopic relations and general theory of interface phenomena for magnetics, we obtain an explicit expression for the surface tension which is in a good agreement with simulation data. The rest of the article is organized as follows. In the next Sec. II we outline the Hubbard-Schofield transformation and derivation of the microscopic expressions for the effective magnetic-like Hamiltonian. In Sec. III we discuss the application of the effective Hamiltonian to the liquid-vapor interface  and compute the surface tension; we also compare the theoretical results with the available simulation data. Finally, in Sec. IV we summarize our findings.

\section{Hubbard-Schofield transformation and magnetic-like Hamiltonian}
\subsection{Hubbard-Schofield transformation}
There is a variety of approaches to perform integral transformations that result in field-theoretical Hamiltonian. We
outline here the derivation of Ref. \cite{hubbard}, which has been further developed in \cite{Brill:98}, making focus
on the derivation detail that will help to understand the nature of the stochastic field. In what follows we will use
the reference system with only repulsive interactions \footnote{In the recent study \cite{Trokhymchuk} the authors
applied the approach of Ref. \cite{Brill:98} for systems with a repulsive and short-range attractive potential.}.

We start from the fluid Hamiltonian $H=H_R+H_A+H_{\rm ex}$:
\begin{equation}
H=\sum_{i<j} v_r ({\bf r}_{ij}) - \sum_{i<j} v({\bf r}_{ij}) +\sum_j g({\bf r}_j), \label{Ham}
\end{equation}
where $v_r(r)$ denotes the  repulsive part of the interaction potential, $-v(r)$ -- the attractive part and $g({\bf
r})$  -- the external potential; ${\bf r}_i$ are the coordinates of $i$-th particle, $i=1, \ldots N$ and ${\bf r}_{ij}=
{\bf r}_i-{\bf r}_j$.  The last two terms of the Hamiltonian (\ref{Ham}) may be written using the Fourier transforms of
the density fluctuations,
$$
n_{\bf k}=\frac{1}{\sqrt{\Omega}} \sum_{j=1}^{N} e^{-i {\bf k} \cdot {\bf r}_j},
$$
of the attractive potential, $v_k=\int v(r)e^{-i {\bf k} \cdot {\bf r}} d {\bf r} $ and of the external potential $g_k=
\Omega^{-1/2} \int g({\bf r})e^{-i {\bf k} \cdot {\bf r}} d {\bf r} $ as
\begin{equation}
-\frac12 \sum_{\bf k } v_k n_{\bf k} n_{-{\bf k}}+ \frac12 v(0)N + \sum_{\bf k } g_{\bf k} n_{-{\bf k}}\, ,
\end{equation}
where $\Omega=L^3$ is the volume of the system, and summation over $k_l={2\pi n_l}/{L}$ with  $l=x,y,z$, and
$n_l=0,\pm 1, \ldots$ is implied. Let $\mu$ be the chemical potential  of the system with the complete Hamiltonian,
(\ref{Ham}), and $\mu_R$ be the chemical potential of the reference system, with the Hamiltonian, $H_R$, which has
only repulsive interactions. If $\left< N \right> = \partial \Xi / \partial \mu $ is the average number of particles in
the system, so that $\rho =  \left< N \right>/\Omega$ is the average number density, we choose the reference system with such chemical
potential $\mu_R$, that the average density $\rho$ is the same in the both systems.

Following Hubbard and Schofield \cite{hubbard} we express  the grand partition function $\Xi(\mu, \Omega, T)$ in terms
of the grand partition function $\Xi_R(\mu_R, \Omega, T)$ of the reference fluid as
\begin{equation}
\label{Xi} \Xi=\Xi_R \left\langle \exp \left\{ \beta \mu' N+\beta \sum_{\bf k }\left[ \frac{v_k}{2} n_{\bf k} n_{-{\bf
k }} -
n_{\bf k} g_{-{\bf k}} \right] \right\} \right\rangle_R \, . \\
\end{equation}
Here $\beta=(k_BT)^{-1}$, with $k_B$ being the Boltzmann constant, $\mu^{\prime} =\mu-\mu_R+\frac12 v(0)$ and
$\left\langle \, \, \, \right\rangle_R$ denotes the average over the reference system with the chemical potential
$\mu_R$. Using the identity:
$$
e^{\frac12 a^2x^2 -bx}=\frac{1}{\sqrt{2 \pi a^2}} \int_{-\infty}^{\infty} e^{-(y+b)^2/(2a^2) + xy}dy
$$
for each ${\bf k}$ in (\ref{Xi}), we obtain after some algebra the ratio $Q=\Xi/\Xi_R$:
\begin{eqnarray}
\label{Q2} Q  \!\! &\propto &  \!\! \int \prod_{\bf k }  d\phi_{\bf k } \left\langle
 \exp \left\{ \sum_{\bf k}
 \phi_{\bf k } n_{-{\bf k} }
\right\} \right\rangle_R \, \, \exp \left\{ \, \frac{\mu^{\prime}}{v_0} \Omega^{1/2}\phi_0 \right\} \nonumber\\
\!\!&\times&  \!\!\exp \left\{ -\frac{1}{2 \beta} \sum_{\bf k} v_k^{-1} \left( \phi_{\bf k }\! +\! \beta g_{\bf k }
\right) \left( \phi_{-{\bf k} } \!+ \!\beta g_{-{\bf k} } \right) \right\}.
\end{eqnarray}
The integration in Eq. (\ref{Q2}) is to be performed under the constraint $\phi_{-{\bf k} }=\phi_{\bf k }^*$, and a
factor which does not affect the subsequent analysis is omitted. Applying the cumulant theorem to the factor
$\left\langle \exp \left\{\sum_{\bf k } \phi_{\bf k }n_{-{\bf k }}\right\}\right\rangle _R$ we arrive at
\cite{hubbard},
\begin{eqnarray}
\label{Q3}
&& Q   \propto \int  \prod_{\bf k }  d\phi_{\bf k } \exp(- \beta {\cal H} ) \\
&&\beta {\cal H}= -\tilde{h} \, \Omega^{1/2} \phi_0
+ \sum_{n=2}^{\infty} \Omega^{1-n/2}
 \sum_{{\bf k}_1, \ldots {\bf k}_n} \tilde{u}_n \,
\phi_{{\bf k}_1} \cdots \phi_{{\bf k}_n} \, , \nonumber
\end{eqnarray}
where the coefficients of the effective magnetic-like Hamiltonian ${\cal H}$ read for $g({\bf r })=0$ \cite{Brill:98}:
\begin{eqnarray}
\label{defhun} && \tilde{h}= \mu^{\prime} v_0^{-1} +\rho \\
 &&\tilde{u}_{2}({\bf k}_1,{\bf k}_2) =\frac{1}{2!} \, \delta_{{\bf k}_1 +{\bf k}_2,0}
\left\{\beta^{-1}v_{k_1}^{-1}- \left\langle n_{{\bf k}_1} n_{-{\bf k}_1} \right\rangle _{cR} \right\}\, ,
\nonumber \\
&&\tilde{u}_{n}\left({\bf k}_1, \ldots {\bf k}_n \right)= -\frac{\Omega^{n/2-1}}{n!} \left\langle n_{{\bf k}_1} \cdots
n_{{\bf k}_n} \right\rangle_{cR} \, \qquad n \ge 3 \,. \nonumber
\end{eqnarray}
Here $\left\langle \, \, \, \right\rangle _{cR}$ denotes the {\it cumulant} average calculated in the (homogeneous)
reference system with density $\rho=\left<N\right>/\Omega$. According to (\ref{Q3}), $Q$ has the form of a partition function of the
system with the field-theoretical Hamiltonian ${\cal H}$, which depends on the order parameter $\phi({\bf r})$
($\phi_{\bf k }$ are  the Fourier components of the order parameter).

Let us analyze the physical meaning of the order parameter $\phi({\bf r})$. From Eq. (\ref{Xi}) directly follows:
$$\frac{\partial \log \Xi }{ \partial g_{-{\bf k}}}= \frac{\partial \log Q }{ \partial g_{-{\bf k}}}=-\beta
\left \langle n_{\bf k} \right \rangle.
$$
On the other hand Eq. (\ref{Q2}) yields for $g \to 0$:
$$\frac{\partial \log Q }{\partial g_{-{\bf k}}}  = -v_k^{-1}\left
\langle \phi_{\bf k} \right \rangle,
$$
where the averaging is to be understood as the integration over all distributions of the order parameter. Thus, we
conclude that $ \langle \phi_{\bf k} \rangle = \beta v_k \langle n_{\bf k} \rangle$, which may be written in terms of
space-dependent field as

\begin{equation}
\label{defopar} \left \langle \phi({\bf r}) \right \rangle = \beta \int \rho ({\bf r}_1) v({\bf r}-{\bf r}_1) d {\bf
r}_1 + \bar{\phi}\,,
\end{equation}
where $\rho({\bf r})=\left \langle n ({\bf r}) \right \rangle $ is the average density and we add an arbitrary
constant, $\bar{\phi}$. This equation suggests the following physical interpretation of the order parameter: $\phi({\bf
r})$ gives the microscopic one-body molecular potential at point ${\bf r}$ (in units of $k_BT=\beta^{-1}$) emerging due
to the attractive part of interaction potential $-v(r)$ from particles distributed in space with microscopic density $n
({\bf r}) =\sum_i \delta ({\bf r}-{\bf r}_i)$. Eq. \eqref{defopar} relates the average quantities. This microscopic
potential is associated with the microscopic force acting on a particle, which may be written for the average values as
$$\beta \left \langle f(\bold{r}) \right \rangle = \nabla \left \langle\phi(\bold{r})\right \rangle = \beta\int \nabla \rho ({\bf r}-{\bf r}_1) v({\bf r}_1) d {\bf r}_1.$$
This force is zero in a uniform system with $\rho({\bf r})=const$ and is directed along the density gradient for
non-uniform systems. In particular, such force arises at an interface, pulling the molecules towards a more dense
phase, thus manifesting the interphase surface tension.  This illustrates that the stochastic field that formally
appears in the KSSHE and HS transformations has a clear physical meaning.

\subsection{Microscopic expressions for the coefficients of magnetic Hamiltonian}

As it is seen  from Eq. (\ref{defhun}) the coefficients  of ${\cal H}$ depend on the correlation functions of the
reference fluid having only repulsive interactions. Using the definition of $l$-particle correlation functions
$g_l({\bf r}_1, \ldots {\bf r}_l)$ \cite{gray,grayA},  one can express the cumulant averages $\left\langle n_{{\bf
k}_1} \cdots n_{{\bf k}_n} \right\rangle _{cR}$, and thus the coefficients $\tilde{u}_{n}\left({\bf k}_1, \ldots {\bf
k}_n \right)$ in terms of the Fourier transforms of $g_l$. Actually, $\tilde{u}_{n}$ depend on the {\it connected}
correlation functions $h_1,h_2, \ldots h_n$, defined as \cite{Brill:98},
\begin{eqnarray} \label{h1h2h3}
&&h_1({\bf r}_1) = \delta({\bf r}_1),  \\
&&h_2({\bf r}_1,{\bf r}_2) =  g_2({\bf r}_1,{\bf r}_2)-1,  \nonumber \\
&&h_3({\bf r}_1,{\bf r}_2,{\bf r}_3) =  g_3({\bf r}_1,{\bf r}_2,{\bf r}_3)- g_2({\bf r}_1,{\bf r}_2) \\
&&~~~~~~~~~~~~~~~~~~~~~~~~-g_2({\bf r}_1,{\bf r}_3)-g_2({\bf r}_2,{\bf r}_3)+2 \,, \nonumber
\end {eqnarray}
etc. For instance, $\tilde{u}_{2}({\bf k}_1,{\bf k}_2)$ depends on $\tilde{h}_2({\bf k}_1)$ ($\tilde{h}_l$ is the
Fourier transforms of $h_l$) as
\begin{equation}
\label{coef1} \tilde{u}_{2} = \frac{1}{2!} \left[(\beta v_k)^{-1}- \rho(1+\rho \tilde{h}_2({\bf k}_1) \right]
\delta_{{\bf k}_1+{\bf k}_2,0} \, .
\end{equation}
Similarly, $\tilde{u}_{3}$ depends on $\tilde{h}_2({\bf k}_{1/2/3})$ and $\tilde{h}_3({\bf k}_1, {\bf k}_2,{\bf k}_3)$,
and $\tilde{u}_4$ depends on $\tilde{h}_2$, $\tilde{h}_3$ and $\tilde{h}_4$, and so on \cite{Brill:98}.

For the subsequent analysis it is instructive to use in the effective Hamiltonian the space-dependent order parameter
$\phi({\bf r})$, instead of its Fourier components $\phi_{\bf k }$. Writing ${\cal H}$ in terms of $\phi({\bf r})$, we
assume that  $\phi({\bf r})$ varies smoothly in space and make the gradient expansion. This corresponds to small ${\bf
k}$ expansion of the coefficients $\tilde{u}_{n}\left({\bf k}_1, \ldots {\bf k}_n \right)$. We keep only the
 square-order gradient terms $\sim (\nabla \phi )^2$ which correspond to $\sim k^2 \phi_k \phi_{-k} $ and omit high-order
gradient terms and cross-terms $\sim (\nabla \phi )^2 \phi^k$ with  $k>0$. In the square gradient approximation
$\tilde{u}_2$ should be expanded as $\tilde{u}_2 =\tilde{u}_2(0)- \tilde{u}_2^{\prime\prime}(0)k^2 +\cdots$, since
$(\nabla \phi)^2 \sim k^2 \phi_{\bf k } \phi_{-{\bf k} }$. The other coefficients $\tilde{u}_n$, where  $n \ge 3$ are
to be taken at zero wave-vectors, as $\tilde{u}_{n}\left(0, 0, \ldots 0 \right)$, since the terms $\sim (\nabla \phi
)^2 \phi^k$ should be omitted. Thus, as it follows from Eq.~(\ref{coef1}) and the discussion below (\ref{coef1}), only
$\tilde{h}^{\prime\prime}_2(0)$ and $\tilde{h}_l({\bf 0}) \equiv \tilde{h}_l(0,0,\ldots 0)$, with $l \ge 2$, are
needed. Using the expansions $v_k=v_0 -v_0^{\prime \prime} k^2 + \ldots$ and $\tilde {h}_2(k) = \tilde{h}_2(0)-
\tilde{h}_2^{\prime \prime} (0) k^2 +\ldots $ (the functions $v_k$ and $\tilde{h}_2$ are even) we obtain for the
coefficients:

\begin{eqnarray}
\label{u2u3u4} \tilde{u}_2 \!\!&=&  \!\!\left[ \!\frac{k_BT}{v_0} \!- \!\rho \!-\! \rho^2  \tilde{h}_2({\bf 0})
\!+\!k^2\left(
\frac{k_BT}{v_0^2}v_0^{\prime \prime} \!-\!\rho^2 \tilde{h}_2^{\prime \prime} ({\bf 0}) \right) \right] \delta_{1,2}  \nonumber \\
 \tilde{u}_3
\!\!&=&\!\! -  \rho\left[1+3 \rho \tilde{h}_2({\bf 0}) +\rho^2 \tilde{h}_3({\bf 0}) \right] \delta_{1,2,3}\\
\tilde{u}_4 \!\!&=&\!\! -  \rho\left[1+7 \rho \tilde{h}_2({\bf
0}) +6\rho^2 \tilde{h}_3({\bf 0}) +\rho^3  \tilde{h}_4({\bf
0}) \right] \delta_{1,2,3,4},  \nonumber
\end{eqnarray}
where we apply the shorthand notation $\delta_{1,2,\ldots ,n,0} \equiv \delta_{{\bf k}_1+{\bf k}_2+ \ldots {\bf
k}_n,0}/n!$. In what follows we will use the relation for  the isothermal compressibility  $\chi_R = \rho^{-1}
(\partial \rho/\partial P_R)_T$ of  the reference fluid ($P_R$ is the pressure of the reference fluid),
\begin{equation} \label{z0chi}
1+ \rho \tilde{h}_2(0) =\rho k_BT \chi_R \equiv z_0.
\end{equation}
We will also use the general relation between the
successive $l$-particle correlation function $g_l({\bf r}_1, \ldots {\bf r}_l)$ \cite{gray},
$$
\chi \rho^2 \frac{\partial}{ \partial \rho} \rho^l g_l = \beta \rho^l \left[ l g_l +\rho \int d{\bf r}_{l+1} (g_{l+1} -
g_l ) \right].
$$
With Eqs. \eqref{h1h2h3} one can express the $l$-particle correlation functions $g_l$ in terms of the {\em connected}
correlation functions $h_l$. Applying the Fourier transform to the resulting equations for $h_l$, we finally arrive at
the following relation for the Fourier  transforms of the functions $\tilde{h}_l$, taken at zero wave vectors ${\bf
k}_1={\bf k}_2=\ldots {\bf k}_l=0$ \cite{Brill:98}:
\begin{equation}
\label{relclus} z_0 \, \rho \frac{ \partial}{\partial \rho} \rho^l \tilde{h}_l({\bf 0}) = \rho^l \left[ l\,
\tilde{h}_l({\bf 0}) + \tilde{h}_{l+1} ({\bf 0}) \right] \, .
\end{equation}
The equation  \eqref{relclus} will be applied for the reference system, where $z_0$ as defined by Eq. \eqref{z0chi}, is
the reduced compressibility of the reference fluid.

Eq.~\eqref{relclus} allows to express $\tilde{h}_{l+1} ({\bf 0})$ in terms of $\tilde{h}_{l} ({\bf 0})$ and its density
derivative. Using this equation iteratively, along with $\tilde{h}_1({\bf 0})=1$, one can express all functions
$\tilde{h}_l({\bf 0})$ in terms of the reduced compressibility $z_0$ and its density derivatives. With
Eqs.~\eqref{u2u3u4} we obtain for the coefficients of the effective Hamiltonian:
\begin{eqnarray}
\label{u2u3u4z0} \tilde{u}_3\!\! &=& \!\!-\rho z_0 (z_0 +z_1) \delta_{1,2,3} \equiv u_3^{\prime} \, \delta_{1,2,3} \\
\tilde{u}_4\!\! &=& \!\! - \rho z_0\left[z_1^2 +z_0(z_0+4z_1+z_2) \right] \delta_{1,2,3,4} \equiv u_4^{\prime}\,
\delta_{1,2,3,4}\, ,\nonumber
\end{eqnarray}
where $z_1= \rho (\partial z_0/\partial \rho)$ and $z_2= \rho^2 (\partial^2 z_0/\partial \rho^2)$. Similarly, one can
obtain all coefficients $\tilde{u}_n$ of the magnetic-like Hamiltonian.

For a reference system with only repulsive interactions one can use the hard--sphere fluid with an appropriately chosen
diameter~\cite{gray,grayA}. For soft (not impulsive) repulsive forces a simple Barker-Henderson relation~\cite{grayA}
\begin{equation}
\label{d}  d=\int_0^{R} \left[ 1- \exp \left( - \beta v_r(r)\right) \right] dr
\end{equation}
gives the effective diameter of the hard-sphere system, corresponding to a repulsive potential $v_r(r)$ vanishing at $r
\ge R$. The fairly accurate Carnahan-Starling equation of state for this system~\cite{gray,grayA} yields for the
reduced compressibility
\begin{equation}
\label{z0} z_0=\left(1-\eta \right)^4/ \left( 1+4\eta +4\eta^2 -4\eta^3 +\eta^4 \right)
\end{equation}
with the packing fraction $\eta={\pi d^3 \rho}/{6}$. For the hard-sphere reference system one can also find
$\tilde{h}^{\prime\prime}_2(0)$. This may be done expressing $\tilde{h}_2(k)$ in terms of the direct correlation
function $\tilde{c}_2(k)$, as $\tilde{h}_2(k)=\tilde{c}_2(k)/\left[1-\rho \tilde{c}_2(k) \right]$ ~\cite{gray,grayA}
 and expanding $\tilde{c}_2(k)$ as $\tilde{c}_2(k)=\tilde{c}_2(0)-\tilde{c}^{\prime\prime}_2(0)k^2+\cdots$,
\begin{equation}
\label{hz0}  \tilde{h}_2(k) = \tilde{h}_2(0) -z_0^2 \tilde{c}^{\prime\prime}_2(0)k^2 +\ldots,
\end{equation}
where we use Eq. \eqref{z0chi} for $\tilde{h}_2(0)$. Hence $\tilde{h}^{\prime\prime}_2(0)= z_0^2
\tilde{c}^{\prime\prime}_2(0)$. The value of $\tilde{c}^{\prime\prime}_2(0)$ may be found from the the Wertheim-Thiele
solution for the direct correlation function of a hard sphere fluid \cite{gray,grayA}:

\begin{equation}
\label{c2pp}\tilde{c}^{\prime\prime}_2(0)= \frac{\pi d^5}{120} \, \frac{\left(16-11\eta +4 \eta^2 \right)}{\left(
1-\eta \right)^{4}}.
\end{equation}

Substituting $\tilde{h}^{\prime\prime}_2(0)$, expressed through $\tilde{c}^{\prime\prime}_2(0)$ from Eq.~\eqref{c2pp},
into Eq.~\eqref{u2u3u4} we recast $\tilde{u}_2$ into the form:

\begin{eqnarray}
\label{u2c2}\tilde{u}_2&=&(a_2^{\prime} +b_2^{\prime} k^2) \delta_{1,2} \\
a_2^{\prime}&=&(\beta v_0)^{-1} -\rho z_0 \nonumber \\
b_2^{\prime}&=& (\beta v_0)^{-1} (v_0^{\prime \prime}/v_0) + \rho^2 z_0^2 \tilde{c}^{\prime\prime}_2(0).  \nonumber
\end{eqnarray}
Now we perform a transformation from the variables $\phi_{\bf k}$ to the space-dependent field $\phi ({\bf r})$. Under
this transformation the integration over the set $\{ \phi_{\bf k} \}$ in Eq.~\eqref{Q3} converts into integration over
the field $\phi ({\bf r})$ and the term $\sim k^2 \phi_{\bf k}\phi_{\bf -k}$ transforms into $ \sim (\nabla \phi_{\bf
k})^2$. As the result we obtain,
\begin{equation}
\label{betaH} \beta {\cal H}[\phi] = \int d{\bf r} \left[ \frac12 \kappa (\nabla \phi)^2  + W(\phi) \right] \, ,
\end{equation}
where
\begin{equation}
\label{vphi1} W(\phi)\!\!=\!\!- h^{\prime} \phi ({\bf r})\!+\! \frac{a_2^{\prime}}{2!}  \phi^2({\bf r}) \!+\!
\frac{u_3^{\prime}}{3!}  \phi^3({\bf r}) \!+ \!\frac{u_4^{\prime}}{4!}  \phi^4({\bf r}) + \ldots
\end{equation}
and we keep only terms up to the fourth order in $\phi({\bf r})$. In Eq.~\eqref{vphi1} $h^{\prime}=\tilde{h}$ is
defined by Eq.~ \eqref{defhun}, $a_2^{\prime}$ by Eq.~\eqref{u2c2} and   $u_3^{\prime}$ and  $u_4^{\prime}$ by
Eqs.~\eqref{u2u3u4z0}. The coefficient at the gradient term reads,
\begin{equation}
\label{kappa} \kappa=\frac{3}{40 \pi d} \left[ \frac{\lambda_{\rm eff}^2}{\beta \epsilon_{\rm eff}}-B \right]
\end{equation}
where $B=4 \eta^2(1-\eta)^4(16-11\eta+4 \eta^2)/(1+4\eta+4\eta^2-4\eta^3+\eta^4)^2$ and  the constants $\epsilon_{\rm
eff}$ and $\lambda_{\rm eff}$ characterize the effective  depth and effective width of the attractive potential $v(r)$:
\begin{eqnarray}
\label{epsef} \epsilon_{\rm eff} &=& \frac{3}{4\pi d^3}\int v(r) d{\bf r} \\
\label{lamef} \lambda_{\rm eff}^2 &=&\frac{5}{3v_0d^2}\int v(r) r^2 d{\bf r}.
\end{eqnarray}
To obtain $\kappa$ we use Eqs.~\eqref{u2c2} and \eqref{c2pp} for $b_2^{\prime}$ and $\tilde{c}^{\prime\prime}_2(0)$
respectively.

The cubic term in the potential $W(\phi)$ may be removed by the shift of the field $\phi \to \phi + \bar{\phi}$, with
the constant field $\bar{\phi}$, chosen to make the term $\sim \phi^3$ vanish. This results in the celebrated
Landau-Ginzburg-Wilson (LGW) Hamiltonian \eqref{betaH} with
\begin{equation}
\label{LGWdef} V(\phi)=- h \phi ({\bf r})+ \frac{a_2}{2!}  \phi^2({\bf r}) + \frac{u_4}{4!} \phi^4 ({\bf r}) \, ,
\end{equation}
and re-normalized coefficients:
\begin{eqnarray}
\label{coefinal}
&&u_4=-\rho z_0 [z_1^2+z_0(z_0+4z_1+z_2)] \nonumber\\
&&a_2=(\beta v_0)^{-1}-\rho\left[z_0+z_3^2/(2\rho u_4)\right] \\
&&h=\mu^{\prime} v_0^{-1} + \left[a_2+z_3^2/(6u_4)\right] (z_3/u_4)   +\rho \, , \nonumber
\end{eqnarray}
where $z_0$, $z_1$ and $z_2$ have been defined above and  $z_3 \equiv - \rho z_0 (z_0+z_1)$. The coefficient $\kappa$
is not affected by the field transformation.

The free energy of the system $F$ with the LGW Hamiltonian may be written in terms of the functional integral over the
statistical field as
\begin{equation}
\label{FH} \beta F = -\log \left( \int {\cal D}\left[ \phi({\bf r}) \right] e^{-\beta H (\phi)} \right),
\end{equation}
where $H \left( \phi ({\bf r}) \right)$ is given by Eq. \eqref{betaH} and  ${\cal D}[ \phi ]$ denotes the functional
(field) integration. For brevity we skip in \eqref{FH} the normalization constant.

\section{Surface tension of liquid-vapor interface}

To illustrate some practical application of  our approach we calculate the surface tension of the liquid-vapor
interface within the mean field (MF) approximation. In the MF approximation only the extremal field $\phi^*({\bf r })$,
which minimizes the free energy, $\delta F[\phi^*({\bf r })]/ \delta \phi({\bf r })=0 $, is taken into account. Using
Eq. \eqref{FH} we obtain for the mean field free energy (see also \cite{Igna:2000,BrezinFeng:84}),
$$
F_{\rm mf}\!=\!{\cal H}[\phi^*]\! =\!\int \! d{\bf r} \left[\frac{\kappa}{2} (\nabla \phi^*)^2 \! + \!V(\phi^*) \right]
\!\!=\!\!\int \! d{\bf r} f(\phi^*,\!\nabla{\phi^*}),
$$
where $f(\phi^*,\nabla{\phi^*}) = \kappa (\nabla \phi^*)^2/2 + V(\phi^*)$ is the free energy density for a general
geometry.

For a flat interface with $\nabla =d/dx$, the equation for the extremal field reads  \cite{Igna:2000,BrezinFeng:84}
\begin{equation}
\label{extfiel} \kappa \frac{d^2 \phi^* }{ d x^2}
= \frac{d V(\phi^*)}{d \phi^*}  \, .
\end{equation}

In the bulk of the two phases, i.e. far  from the surface, the order parameter
takes constant values, $\phi^*_1$ at $x \to -\infty$, and
 $\phi^*_2$ at $x \to \infty$, which are related to the mean densities of
these phases -- of  the liquid, $\rho_l$, and of the vapor, $\rho_g$ density respectively.
As stated above and follows from Eq. \eqref{defopar}, the extremal fields $\phi^*_{1,2}$ in the bulk of the phases are linearly related to the densities of the phases,   $\phi^*_{1,2}=\beta v_0 \rho_{l,g}+\bar{\phi}$.  Hence the standard phase equilibrium conditions for the free energy density, $f^{\prime} (\rho_l)= f^{\prime} (\rho_g)$ and  $f(\rho_l)+ \rho_l f^{\prime} (\rho_l)= f(\rho_g)+ \rho_gf^{\prime} (\rho_g)$ for two bulk phases may be written as
\begin{eqnarray*}
&&V^{\prime}(\phi^*_1)=V^{\prime}(\phi^*_2)\\
&&V(\phi^*_1)+\phi^*_1 V^{\prime}(\phi^*_1)=
V^{\prime}(\phi^*_2)+\phi^*_2 V^{\prime}(\phi^*_2)
\end{eqnarray*}
which is the double-tangent construction for the fields   $\phi^*_1$ and $\phi^*_2$.

If we choose the interface located at $x=0$, the first integral of Eq. (\ref{extfiel}) yields:
\begin{equation}
\label{fint} \frac12 \kappa \left(\frac{d \phi^*}{d x}\right)^2=
\begin{cases}
V(\phi^*)-V(\phi^*_1) &  x \leq 0 \\
V(\phi^*)-V(\phi^*_2) & x >0 \, .
\end{cases}
\end{equation}
The surface tension $\gamma$ is equal to the difference per unit area between
the  free energy, calculated for the space-dependent $\phi^*({\bf r })$ and
that for $\phi^*_1$ for $x<0$ and $\phi^*_2$ for $x>0$. If the
order parameter at the interface equals $\phi^*_0$, which may be
chosen from the condition $\phi^*_1<\phi^*_0< \phi^*_2$,
$V^{\prime}(\phi^*_0)=0$, straightforward calculations yield for the surface
tension with $V_{1,2}= V(\phi^*_{1,2})$
(see also \cite{Igna:2000,BrezinFeng:84}):
\begin{equation}
\label{surt1}
\beta \gamma \!=\! \int_{\phi^*_1}^{\phi^*_0}
\sqrt{2\kappa[V(\phi)-V_1]}d\phi
+\int_{\phi^*_0}^{\phi^*_2}
\sqrt{2\kappa[V(\phi)-V_2]}d\phi.
\end{equation}
Now we choose the system for which
the coefficient $h$ in (\ref{coefinal}) vanishes, that is,
$V=\frac12a_2 \phi^{*2}+\frac{1}{4!}u_4\phi^{*4}$. For this system
$\phi^*_{1,2}=\pm (-6a_2/u_4)^{1/2}$, $\phi^*_{0}=0$, and the
solution to  Eq.(\ref{extfiel}) reads,
$$\phi^*(x)=(-6a_2/u_4)^{1/2}\tanh(x/\xi_0)$$
with the interface width $\xi_0 = (-\kappa/2a_2)^{1/2}$
\cite{Igna:2000,BrezinFeng:84}. The solution is symmetric and has zero volume average, $\bar{\phi^*} \equiv \Omega^{-1}\int \phi^*({\bf r})d{\bf r}=0$.

Averaging Eq. (\ref{defopar}) over the volume  yields
$\bar{\phi^*}=\beta \bar{\rho}v_0 + \bar{\phi}=0$, implying that
$\bar{\phi}=-\beta v_0 \bar{\rho}$, where
$\bar{\rho}=\Omega^{-1}\int \rho({\bf r})d{\bf r}= N/\Omega$ is the
averaged over the volume density. Since $\phi^*_{1}=-\phi^*_{2}$ and
simultaneously $\phi^*_{1,2}=\pm \beta v_0 \rho_{l,g}+\bar{\phi}$, we
conclude that $\bar{\rho}= (\rho_l+\rho_g)/2$, i.e. that the
averaged density of our
system is the mean between the liquid and vapor density. Naturally, this
is the density of our homogeneous reference
system, with the same volume and number of particles.
With the above values of $\phi^*_{1,2}$ and $\phi^*_{0}$,  the integration
in (\ref{surt1}) is easily performed yielding:
\begin{equation}
\gamma/k_BT  =4 \left( - 2 \kappa a_2^3/u_4^2 \right)^{1/2} \, ,
\label{surt2}
\end{equation}
where microscopic expressions for the constants  $a_2$, $u_4$ and  $\kappa$, are given by Eqs.\eqref{kappa} and \eqref{coefinal} where the density ${\rho}= (\rho_l+\rho_g)/2$ of the reference fluid is to be used.

Not far from the critical point ($\rho_c$, $T_c$), one can
approximate, $(\rho_l+\rho_g)/2 \simeq \rho_c$ and thus use
 $\rho_c$ as the reference density. In particular
one can write for $a_2$:
$a_2 \simeq a_2(\beta, \rho_c)=(\beta v_0)^{-1}-\rho_c[z_0+z_3^2/(2\rho u_4)]_c$
(see (\ref{coefinal})). If we then use the  the mean field
condition for the critical point, $a_2(\beta_c, \rho_c)=0$ \cite{Amit:78},
we obtain
$a_2=(\beta v_0)^{-1}-(\beta_c v_0)^{-1}=-\alpha \tau$,
and finally for the surface tension:
\begin{equation}
\frac{\gamma}{k_BT} = 4 \left( \frac {2 \kappa_c \alpha^3}{u_{4c}^2}  \right)^{1/2}
\tau^{3/2}\, ,
\label{surt3}
\end{equation}
where $\alpha=(\beta_c v_0)^{-1}$, $\tau=(T_c-T)/T_c$, and the coefficients $u_{4,c}$ and $\kappa_c$ are to be
calculated at $\rho=\rho_c$, $T=T_c$. Eq. \eqref{surt3} is the main result of the present study. It gives an explicit
{\em analytical} expression for the surface tension in terms of temperature,  density and parameters of the interaction
potential. It is worth noting that Eq. \eqref{surt3} demonstrates (as expected for the mean-field analysis),  the
classical critical exponent $3/2$, that is, $\gamma \sim \tau^{3/2}$, as was firstly observed by Widom
\cite{widom1965surface,landau2013course}.

\begin{figure}
\centering
\includegraphics[width=8cm]{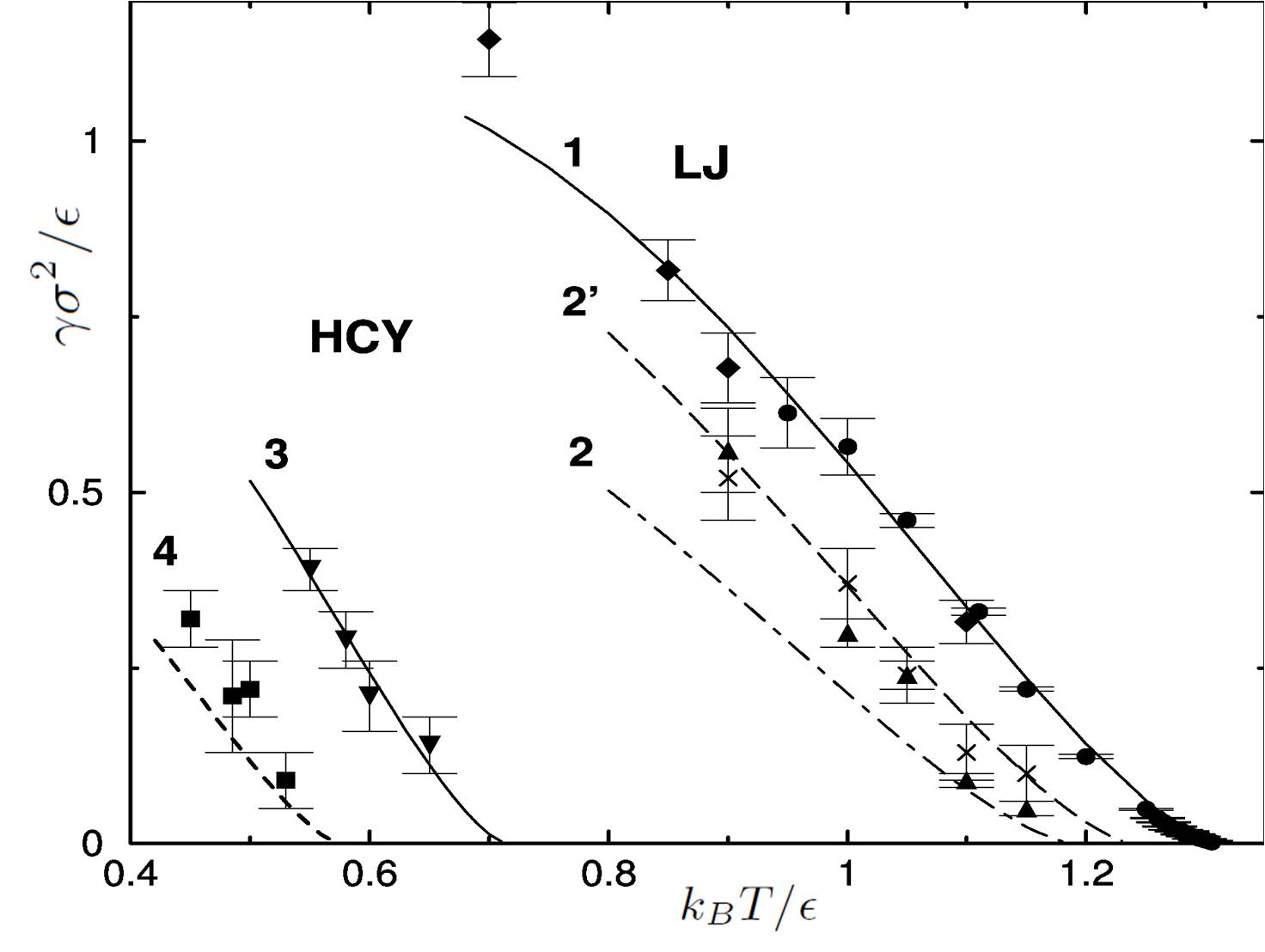}
\caption{Reduced surface tension $\gamma \sigma^2/\epsilon$ as a function of the reduced temperature $k_BT/\epsilon$.
Curves -- theory, Eq. (\ref{surt3}), points -- numerical data.  LJ-fluid: curve $1$, diamonds \cite{MDsur,MDsurA} and
circles \cite{Panagiotopoulos:2001}. HCY-fluid: curves $2$, $2^{\prime}$, $3$ and $4$, stars and triangles up
($\lambda=1.8$ \cite{Trokhymchuk:2001}), triangles down ($\lambda=3.0$, \cite{Trokhymchuk:2001}) and squares
($\lambda=4.0$, \cite{Trokhymchuk:2001}). Critical parameters and $\sigma$, $\epsilon$, $\lambda$ are taken from Refs.
\cite{Panagiotopoulos:2001,Lomba:1994,Duh:1997,Trokhymchuk:2001}, see the text for detail.} \label{fig:surten}
\end{figure}

The theoretical predictions, Eq. (\ref{surt3}), have been compared with the available data of numerical experiments for
the Lennard-Jones (LJ) and hard-core Yukawa (HCY) fluids. For these systems the standard WCA partition (see e.g.
\cite{grayA}) of the potential into attractive and repulsive parts has  been applied \cite{Brill:98}. The numerical
data have been obtained for the LJ-fluid by means of molecular dynamics (MD)  \cite{MDsur,MDsurA} and Monte Carlo
\cite{Panagiotopoulos:2001}. For the HCY-fluid the MC and MD \cite{Trokhymchuk:2001} have been also applied. The
critical parameters for the LJ-fluid were taken from Ref. \cite{Panagiotopoulos:2001}. For the HCY fluid we used the
critical parameters from  Ref. \cite{Lomba:1994} for the curves $2$, $3$, $4$, and parameters from Ref. \cite{Duh:1997}
for the curve $2^{\prime}$. The values of $\sigma$ and  $\epsilon$ for the LJ potential were taken from Ref.
\cite{Panagiotopoulos:2001} and $\sigma$, $\epsilon$, $\lambda$ for the HCY potential from Ref.
\cite{Trokhymchuk:2001}.

As follows from Fig. \ref{fig:surten} our theory is in a good agreement with the numerical experiments. It is expected,
however, that the agreement would be worse in the very close vicinity of the critical point, where the mean field
theory loses its accuracy. The accuracy of numerical simulations also decreases in the vicinity of the critical point
\cite{BrillVal1998}. Eq. (\ref{surt3}) is quite sensitive to the critical parameters $\rho_c$, $T_c$. While these are
known rather accurately for the LJ fluid, they are estimated with much larger uncertainty for the HCY fluid. This is
demonstrated in Fig. \ref{fig:surten}, where two theoretical curves ($2$ and $2^{\prime}$) correspond to the same HCY
fluid but with $\rho_c$, $T_c$, taken from different sources ($\rho_c$ and $T_c$ differ by about $4\%$).

\section{Conclusion}
We develop a theory of inhomogeneous simple fluids based on the
microscopic one-body potential in fluid, which naturally emerges in the Hubbard-Schofield (HS) transformation. We demonstrate that the "technical" field variable $\phi({\bf r})$, associated with the HS transformation, possesses a clear physical meaning. It gives the molecular potential at point ${\bf r}$ (in units of $k_BT$) from the attractive part of the inter-particle potential of molecules located in the vicinity of ${\bf r}$. Hence  $\phi({\bf r})$ depends on both -- on the particle density $\rho({\bf r})$ and on the attractive potential $v(r)$, being the convolution of $\rho({\bf r})$ and $v({\bf r})$. As the result, the microscopic field $\phi({\bf r})$ varies much more smoothly, even in the interface region than the local density $\rho({\bf r})$ itself. The smooth variation of $\phi({\bf r})$ guarantees the accuracy of the small gradient expansion, applied for the field-dependent Hamiltonian. Moreover, any additional smoothing procedure is not required. This makes the approach more simple and presumably more reliable. In contrast, the density functional theory, based on the local density, see e.g. \cite{henderson1992fundamentals}, exploits the smoothing of $\rho({\bf r})$ due to its sharp variation at the  interface. The smoothing weight  function is commonly chosen {\it ad hoc}, see e.g.    \cite{henderson1992fundamentals,Forstman:97,Forstman:97a}.

Using the microscopic molecular field approach, which steams from the HS transformation we calculate the surface
tension $\gamma$ for the liquid-vapor interface. Here we apply the mean-field approximation which considers only
average molecular field and ignores the field fluctuations. We obtain an explicit analytical result for $\gamma$, which
expresses this quantity in terms of temperature and density of the system and parameters of the inter-molecular
potential. The theoretical predictions  for the surface tension are in a good agreement with the results of numerical
experiments. The mean-field approach loses however its accuracy in the very close vicinity to the critical point, where
the near-critical fluctuations become important. The account of the critical fluctuation for $\gamma$ is
straightforward and may be done applying the technique developed in Ref. \cite{BrezinFeng:84}. Up to our knowledge, our
study reports for the first time simple analytical expression for the surface tension that agrees well with the
numerical experiments.

\section{Acknowledgements}
NB gratefully acknowledges the financial support from the Russian Foundation for Basic Research under grant
18-29-19198. YB thanks Russian Foundation for Basic Research Grant No 18-29-06008 for financial support. JMR has been
supported by MICINN of the Spanish Government under Grant No. PGC2018-098373-B-I00 and by the Catalan Goverment under
the Grant 2017-SGR-884


\end{document}